\hfuzz 1pt
\font\titlefont=cmbx10 scaled\magstep1
\magnification=\magstep1

\null
\vskip 1.5cm
\centerline{\titlefont EXPERIMENTAL LIMITS ON COMPLETE POSITIVITY}
\smallskip
\centerline{\titlefont FROM THE K-$\overline{\hbox{K}}$ SYSTEM}
\vskip 2.5cm
\centerline{\bf F. Benatti}
\smallskip
\centerline{Dipartimento di Fisica Teorica, Universit\`a di Trieste}
\centerline{Strada Costiera 11, 34014 Trieste, Italy}
\centerline{and}
\centerline{Istituto Nazionale di Fisica Nucleare, Sezione di 
Trieste}
\vskip 1cm
\centerline{\bf R. Floreanini}
\smallskip
\centerline{Istituto Nazionale di Fisica Nucleare, Sezione di 
Trieste}
\centerline{Dipartimento di Fisica Teorica, Universit\`a di Trieste}
\centerline{Strada Costiera 11, 34014 Trieste, Italy}
\vskip 2cm
\centerline{\bf Abstract}
\smallskip
\midinsert
\narrower\narrower\noindent
Available data on measured observables allow deriving estimates
for some of the phenomenological parameters 
that characterize the time-evolution
and decay of the neutral kaon system based on the hypothesis of complete
positivity. The present experimental uncertainties are still too large
to permit a full test of complete positivity.
\endinsert
\bigskip
\vfil\eject

Completely positive evolution maps are the most natural
tools for the description of the time evolution of open quantum 
dynamical systems.[1, 2] These systems can be modeled as being small
subsystems in interaction with suitable large environments.
The global evolution of the closed compound system is described by an
unitary map, while the reduced dynamics of the subsystem usually
develops dissipation and irreversibility.

Assuming a weak coupling between subsystem and environment, the reduced
time-evolution is free from non-linear feedback or memory effects;
moreover, it possesses very basic and fundamental properties, like
forward in time composition (semigroup property), probability conservation
and entropy increase.

On the states of the subsystem, conveniently described by density matrices,
the reduced evolution is realized by a one-parameter (= time) family of
maps, transforming density matrices into density matrices and with the
additional important property of being completely positive.[3-6] This set
of transformations forms a so-called dynamical semigroup.

Decay systems can be viewed as specific examples of open quantum systems
and therefore dynamical semigroups can be effectively used in their description.
Indeed, the time evolution of the neutral kaon system have been recently
studied within this frame- work.[7, 8] In practice, 
assuming a complete positive
map being at the basis of the dynamics of the kaons 
results in the addition to the
standard Wiesskopf-Wigner evolution of an extra linear piece,
characterized by six real phenomenological parameters. In the following
we shall explain in detail how bounds on some of these parameters
can be obtained from available experimental data.

\bigskip

As usual, we shall model the evolution and decay of 
the $K^0$-$\overline{K^0}$ system
by means of a two-dimensional Hilbert space. [9]
A convenient orthonormal basis in this space is given by the $CP$-eigenstates 
$|K_1\rangle$ and $|K_2\rangle$:
$$
|K_1\rangle={1\over\sqrt{2}}\Big[|K^0\rangle+|\overline{K^0}\rangle\Big]\ ,\quad
|K_2\rangle={1\over\sqrt{2}}\Big[|K^0\rangle-|\overline{K^0}\rangle\Big]\ .
\eqno(1)
$$
As already mentioned, states of a quantum system evolving in time can be
suitably described by a density matrix $\rho$.
This is a positive hermitian operator, {\it i.e.} with positive eigenvalues,
and constant trace (for unitary evolutions). 
In the case of the kaon system, $\rho$ is a two-dimensional
matrix. With respect to the basis (1) it can be written as:
$$
\rho=\left(\matrix{
\rho_1&\rho_3\cr
\rho_4&\rho_2}\right)\ , \eqno(2)
$$
where $\rho_4\equiv\rho_3^*$, and $*$ signifies complex conjugation.

The evolution in time of this matrix can be described in general by an equation
of the following form:
$$
{\partial\rho(t)\over\partial t}=-iH\, \rho(t)+i\rho(t)\, H^\dagger 
+L[\rho] .\eqno(3)
$$
The first two pieces in the r.h.s. constitute the standard Weisskopf-Wigner
contribution, while $L$ is a linear map that is fully determined 
by the condition of complete positivity.

The effective hamiltonian $H$ (the Weisskopf-Wigner hamiltonian)
includes a nonhermitian part, that characterizes 
the natural width of the states:
$$
H=M-{i\over 2}{\mit\Gamma}\ ,\eqno(4)
$$ 
with $M$ and $\mit\Gamma$ hermitian $2\times 2$ matrices.
The entries of these matrices can be expressed in terms of the 
complex parameters $\epsilon_S$, $\epsilon_L$, appearing in the
eigenstates of $H$, 
$$
|K_S\rangle=N_S\left(\matrix{1\cr\epsilon_S}\right)\ ,\quad
|K_L\rangle=N_L\left(\matrix{\epsilon_L\cr 1}\right)\ ,\eqno(5)
$$
and the four real parameters, $m_S$, $\gamma_S$ and $m_L$, $\gamma_L$
characterizing the eigenvalues of $H$: 
$$
\lambda_S=m_S-{i\over 2}\gamma_S\ ,\quad
\lambda_L=m_L-{i\over 2}\gamma_L\ ,\eqno(6)
$$
where $N_S=(1+|\epsilon_S|^2)^{-1/2}$ and $N_L=(1+|\epsilon_L|^2)^{-1/2}$ 
are normalization factors.
It proves convenient to use also the following
positive combinations:
$$
\Delta\Gamma=\gamma_S-\gamma_L\ ,\qquad
\Delta m=m_L-m_S\ ,\eqno(7)
$$
corresponding to the differences between decay widths and masses of the
states $K_S$ and $K_L$, as well as of the complex quantities:
$$
\Gamma_\pm=\Gamma\pm i \Delta m\ ,\qquad 
\Delta\Gamma_\pm=\Delta\Gamma\pm 2i\Delta m\eqno(8)
$$
with $\Gamma=(\gamma_S+\gamma_L)/2$.

The explicit form of the piece $L[\rho]$ can be most simply given by 
expanding $\rho$ in terms of Pauli matrices $\sigma_i$ and the identity
$\sigma_0$: $\rho=\rho_\mu\, \sigma_\mu$, $\mu=\,0$, 1, 2, 3.
In this way, the map $L[\rho]$ can be represented by
a symmetric $4\times 4$ matrix $\big[L_{\mu\nu}\big]$, 
acting on the column vector with components $(\rho_0,\rho_1,\rho_2,\rho_3)$.
It can be parametrized by the six real 
constants $a$, $b$, $c$, $\alpha$, $\beta$, and $\gamma$: [7, 8]
$$
\big[L_{\mu\nu}\big]=-2\left(\matrix{0&0&0&0\cr
                                     0&a&b&c\cr
                                     0&b&\alpha&\beta\cr
                                     0&c&\beta&\gamma\cr}\right)
\ ,\eqno(9)
$$
with $a$, $\alpha$ and $\gamma$ non-negative.
These parameters are not all independent; to assure the complete positivity
of the map $L$, they have to satisfy
the following inequalities:
$$
\eqalign{&a\leq \alpha+\gamma\ ,\phantom{\big(\beta^2\big)^2}\cr
         &\alpha\leq a+\gamma\ ,\phantom{\big(\beta^2\big)^2}\cr
         &\gamma\leq a+\alpha\ ,\phantom{\big(\beta^2\big)^2}\cr}\quad
\eqalign{&4b^2\leq \gamma^2-\big(a-\alpha\big)^2\ ,\cr
         &4c^2\leq \alpha^2-\big(a-\gamma\big)^2\ ,\cr
         &4\beta^2\leq a^2-\big(\alpha-\gamma\big)^2\ .\cr}
\eqno(10)
$$
(Notice that from the above inequalities automatically follows that
the map $L$ is positive; indeed, complete positivity is a stronger
request than simple positivity.)
 
In absence of the additional piece $L[\rho]$, the evolution equation
(3) would transform pure states ({\it i.e.} states of the form 
$|\psi\rangle\langle\psi|$) into pure states, in spite of the fact that
probability is not conserved, since
$\hbox{Tr}\left[\rho(t)]\right]\leq\hbox{Tr}\left[\rho(0)\right]$.
However, this is due to the presence of a non-hermitian part in the effective 
hamiltonian $H$ and not to a mixing-enhancing mechanism
that makes $\rho(t)$ less ordered in time. Instead, loss of quantum 
coherence shows up when the extra piece $L[\rho]$ is also present:
it produces dissipation and possible transitions from pure states to
mixed states.

Equations of the general form (3), but with different $L$, have also been used
to describe the time evolution of the $K^0$-$\overline{K^0}$ system. [10-14, 15]
They are based on the general idea that the quantum fluctuations of the 
gravitational field at Planck's length produces loss of
phase-coherence. [16, 17] In those discussions, 
among other more obvious requirements
like conservation of probability and entropy increase,
the linear map $L[\rho]$ is constrained to be simply positive and not
completely positive.

In the case of the neutral kaons, complete positivity of the time
evolution $\gamma_t$, transforming the initial density matrix $\rho(0)$
into $\rho(t)$, amounts to the positivity of the natural extension of $\gamma_t$
to a linear transformation on the states of a larger system
consisting of the kaon system and another finite-level system of 
arbitrary dimension. 
Simple positivity guarantees that the eigenvalues of any density
matrix for the neutral kaon system remain positive. Complete positivity 
is a stronger property in the sense that the same holds for 
density matrices of the compound system.

Although at first sight the requirement of complete positivity of the
evolution equation (3) instead of the much milder simple positivity seems
a mere technical request, it has far-reaching consequences. 
In particular, when the additional finite-level system alluded above
is taken to be another kaon system (a physical situation commonly
encountered in the so-called $\phi$-factories), the complete positivity of the
time-evolution (3) assures the absence of unphysical effects, like
the appearance of negative probabilities, that could occur for simple
positive dynamics. [18]
   
In the following we shall compare the predictions of the completely
positive dynamics (3), (9) with available experimental data on 
the neutral kaon system. In principle, this allows a complete determination
of the values of the phenomenological parameters appearing in the matrix (9),
and therefore a test on the condition of complete positivity, {\it i.e.}
of the inequalities (10).
In practice, the experimental data are not yet accurate enough to test
the conditions (10); nevertheless, estimates
on the three parameters $c$, $\beta$ and $\gamma$ will be obtained.

The first step is to solve the equation (3) for the density matrix $\rho$
with arbitrary initial conditions. It is convenient to use perturbation
theory, assuming $a$, $b$, $c$, $\alpha$, $\beta$ and $\gamma$ to be small,
of the same order of magnitude of $\epsilon_S \Delta\Gamma$ and
$\epsilon_L \Delta\Gamma$. To first order in the small parameters, the
time-dependence of the four components of the density matrix $\rho(t)$
are explicitly given by: [8]
$$
\eqalignno{
\rho_1(t)=&\, e^{-\gamma_S t}\rho_1+
{\gamma\over\Delta\Gamma}\left(e^{-\gamma_L t}-e^{-\gamma_S t}\right)\rho_2\cr
+&\left(\epsilon_L^*-{2C\over\Delta\Gamma_+}\right)
\left(e^{-\Gamma_- t}-e^{-\gamma_S t}\right)\rho_3
+\left(\epsilon_L-{2C^*\over\Delta\Gamma_-}\right)
\left(e^{-\Gamma_+ t}-e^{-\gamma_S t}\right)\rho_4\ ,&(11a)\cr
&\hskip 12cm\cr
\rho_2(t)=&\, e^{-\gamma_L t}\rho_2+
{\gamma\over\Delta\Gamma}\left(e^{-\gamma_L t}-e^{-\gamma_S t}\right)\rho_1\cr
+&\left(\epsilon_S-{2C\over\Delta\Gamma_-}\right)
\left(e^{-\Gamma_- t}-e^{-\gamma_L t}\right)\rho_3
+\left(\epsilon_S^*-{2C^*\over\Delta\Gamma_+}\right)
\left(e^{-\Gamma_+ t}-e^{-\gamma_L t}\right)\rho_4\ ,&(11b)\cr
&\hskip 4cm\cr
\rho_3(t)=& \, e^{-(\Gamma_-+A-\gamma)t}\rho_3
-{iB\over2\Delta m}\left(e^{-\Gamma_- t}-e^{-\Gamma_+ t}\right)\rho_4\cr
-&\left(\epsilon_S^*+{2C^*\over\Delta\Gamma_+}\right)
\left(e^{-\Gamma_- t}-e^{-\gamma_S t}\right)\rho_1
-\left(\epsilon_L+{2C^*\over\Delta\Gamma_-}\right)
\left(e^{-\Gamma_- t}-e^{-\gamma_L t}\right)\rho_2\ ,&(11c)\cr
&\hskip 4cm\cr
\rho_4(t)=& \, e^{-(\Gamma_++A-\gamma)t}\rho_4
+{iB^*\over2\Delta m}\left(e^{-\Gamma_+ t}-e^{-\Gamma_- t}\right)\rho_3\cr
-&\left(\epsilon_S+{2C\over\Delta\Gamma_-}\right)
\left(e^{-\Gamma_+ t}-e^{-\gamma_S t}\right)\rho_1
-\left(\epsilon_L^*+{2C\over\Delta\Gamma_+}\right)
\left(e^{-\Gamma_+ t}-e^{-\gamma_L t}\right)\rho_2\ ,&(11d)\cr}
$$
where $\rho_1$, $\rho_2$, $\rho_3$ and $\rho_4\equiv\rho_3^*$ are the
initial values at $t=\,0$, and the following convenient notations
have been used:
$$
A=\alpha+a\ ,\quad B=\alpha-a+2ib\ ,\quad C=c+i\beta\ .\eqno(12)
$$

Furthermore, from (3) and (9) 
one can extract the contributions $\rho_L$ and $\rho_S$  
that correspond to the $K_L$ and $K_S$ neutral kaons:
$$
\rho_L=\left(\matrix{
\left|\epsilon_L+{2C^*\over\Delta\Gamma_-}\right|^2+{\gamma\over\Delta\Gamma}
-8\left|{C\over\Delta\Gamma_+}\right|^2
-4\,{\cal R}e\left({\epsilon_L C\over\Delta\Gamma}\right)
& \epsilon_L+{2 C^*\over\Delta\Gamma_-}\cr
\epsilon_L^*+{2 C\over\Delta\Gamma_+} & 1}\right)\ ,\eqno(13a)
$$
and 
$$
\rho_S=\left(\matrix{ 1 & 
\epsilon_S^*+{2 C^*\over\Delta\Gamma_+}\cr
\epsilon_S+{2 C\over\Delta\Gamma_-} &
\left|\epsilon_S+{2C\over\Delta\Gamma_-}\right|^2 -{\gamma\over\Delta\Gamma}
-8\left|{C\over\Delta\Gamma_+}\right|^2
-4\,{\cal R}e\left({\epsilon_S C^*\over\Delta\Gamma}\right)}\right)
\ .\eqno(13b)
$$
In these expressions we have
kept only the leading contributions in each entry of the two matrices;
this approximation will suffice for the subsequent discussions.

The solutions (11) and (13) can now be used to compute certain observables
for the $K^0-\overline{K^0}$ system; indeed, any physical property of the
neutral kaons can be extracted from the density matrix $\rho(t)$
by taking its trace with suitable hermitian operators.

Useful observables are associated with the decays of the neutral
kaons into pion states, or into semileptonic
states $\pi\ell\nu$. 
The amplitudes for the decay of a $K^0$ state into $\pi^+\pi^-$
and $\pi^0\pi^0$ final states are usually parametrized as follows,
in terms of the $s$-wave phase-shifts $\delta_i$ and the complex coefficients
$A_i$, $B_i$, $i=1,\ 2$: [19]
$$
\eqalignno{
&{\cal A}(K^0\rightarrow \pi^+\pi^-)=(A_0+B_0)\, e^{i\delta_0}+{1\over\sqrt2}\,
(A_2+B_2)\, e^{i\delta_2}\ ,&(14a)\cr
&{\cal A}(K^0\rightarrow \pi^0\pi^0)=(A_0+B_0)\, e^{i\delta_0}-\sqrt{2}\,
(A_2+B_2)\, e^{i\delta_2}\ ,&(14b)}
$$
where the indices 0, 2 refers to the total isospin. The amplitudes for
the $\overline{K^0}$ decays are obtained from these with the substitutions:
$A_i\rightarrow A_i^*$ and $B_i\rightarrow -B_i^*$.
The imaginary parts of $A_i$ signals direct $CP$-violation, while a non zero
value for $B_i$ will also break $CPT$ invariance.

To construct the operators that describe these two pion decays in
the formalism of density matrices, one has to pass to the $K_1$, $K_2$
basis of Eq.(1). It is convenient to label the corresponding decay amplitudes
as follows:
$$
\eqalign{
&{\cal A}(K_1\rightarrow \pi^+\pi^-)=X_{+-}\ ,\cr
&{\cal A}(K_1\rightarrow \pi^0\pi^0)=X_{00}\ ,}\qquad\quad
\eqalign{
&{\cal A}(K_2\rightarrow \pi^+\pi^-)=Y_{+-}\ 
{\cal A}(K_1\rightarrow \pi^+\pi^-)\ ,\cr
&{\cal A}(K_2\rightarrow \pi^0\pi^0)=Y_{00}\ 
{\cal A}(K_1\rightarrow \pi^0\pi^0)\ .}
\eqno(15)
$$
The complex parameters $X$ and $Y$ can be easily expressed in terms
of $A_i$, $B_i$ and $\delta_i$ of (14). Taking the phase convention for which
$A_0$ is real, up to first order in the $CP$ and $CPT$ violating parameters,
one explicitly finds:
$$
\eqalignno{
&X_{+-}=\sqrt{2}\, \Big[A_0+i{\cal I}m(B_0)\Big]\, e^{i\delta_0}
+\Big[{\cal R}e(A_2)+i{\cal I}m(B_2)\Big]\, e^{i\delta_2}\ ,&(16a)\cr
&X_{00}=\sqrt{2}\, \Big[A_0+i{\cal I}m(B_0)\Big]\, e^{i\delta_0}
-2\Big[{\cal R}e(A_2)+i{\cal I}m(B_2)\Big]\, e^{i\delta_2}\ ,&(16a)}
$$
while
$$
Y_{+-}={ {\cal R}e(B_0)\over A_0}+\epsilon^\prime\ ,\qquad
Y_{00}={ {\cal R}e(B_0)\over A_0}-2\epsilon^\prime\ ,
\eqno(17)
$$
where
$$
\epsilon^\prime= {e^{i(\delta_2-\delta_0)}\over\sqrt2}\,
{{\cal R}e(A_2)\over A_0}\, \left[
i{{\cal I}m(A_2)\over {\cal R}e(A_2) }
+{{\cal R}e(B_2)\over  {\cal R}e(A_2) }
-{ {\cal R}e(B_0)\over A_0}\right]\ ,\eqno(18)
$$
is the parameter that uniquely signals $CP$ and $CPT$ violation in the
transition amplitudes.

Using (15), the operators that describe the 
$\pi^+\pi^-$ and $\pi^0\pi^0$ final states and include direct $CP$
and $CPT$ violations are readily found: [15]
$$
{\cal O}_{+-}=|X_{+-}|^2\ \left[\matrix{1&Y_{+-}\cr
                              Y_{+-}^*&|Y_{+-}|^2\cr}\right]\ ,\qquad
{\cal O}_{00}=|X_{00}|^2\ \left[\matrix{1&Y_{00}\cr
                              Y_{00}^*&|Y_{00}|^2\cr}\right]\ .
\eqno(19)
$$
With the help of these matrices, one can now compute the time-dependent
decay rate of a neutral kaon into two pions. In the case of charged pions, 
for a generic initial state, one has:
$$
R_{+-}(t)={
{\rm Tr}\Big[\rho(t){\cal O}_{+-}\Big]\over
{\rm Tr}\Big[\rho(0){\cal O}_{+-}\Big]}\ .\eqno(20)
$$
For a state which is initially a pure $K^0$,
the corresponding density matrix evolves in time according
to (3), with the initial condition 
$$
\rho(0)=\rho_{K^0}=
|K^0\rangle\langle K^0|\equiv{1\over2}\left(\matrix{1&1\cr
                              1&1}\right)
\ .\eqno(21)
$$
In this case, the observable (20) takes the general form:
$$
R_{+-}(t)=e^{-\gamma_S t}+R_{+-}^L\, e^{-\gamma_L t}+
2\, e^{-\Gamma t}\, |\eta_{+-}|\cos(\Delta m\, t-\phi_{+-})
\ .\eqno(22)
$$
The two-pion decay rate for the $K_L$ state can be obtained using $(13a)$:
$$
R_{+-}^L=
\left|\epsilon_L+{2 C^*\over\Delta\Gamma_-}+Y_{+-}\right|^2
+{\gamma\over\Delta\Gamma}-8\left|{C\over\Delta\Gamma_+}\right|^2
-4\, {\cal R}e\left({\epsilon_L C\over\Delta\Gamma}\right)\ ,\eqno(23)
$$
while the interference term is determined by the combination
$$
\epsilon_L-{2C^*\over\Delta\Gamma_-}+Y_{+-}
\equiv\eta_{+-}=|\eta_{+-}|\, e^{i\phi_{+-}}\ .\eqno(24)
$$
Similar formulas hold for the decay of a $K^0$ into neutral pions:
one just needs to substitute the index $+-$ with $00$.

Notice that in the standard quantum mechanical case, the $K_L$ two-pion
decay rate $R^L_{+-}$ is quadratic in the parameters $\epsilon_L$
and $Y_{+-}$. To get a consistent expression when also the term
$L[\rho]$ in (3) is present, it is necessary to take into account terms up to
second order in all small quantities, as in (23). This is the only case among
the discussed observables for which second order terms are needed.

The expressions (23) and (24) can be directly compared with 
the experimental data.
However, some care needs to be used in the choice of the
proper measured results.
In fact, in fitting their data on the two pion
decay, the various experimental groups use the standard quantum mechanical
formula for $R_{+-}(t)$, obtained from (22) by setting 
$R_{+-}^L=|\eta_{+-}|^2$. However, as discussed in [15], different experimental
setups measured the time dependence of $R_{+-}(t)$ focusing either into the
interference region or the long-time tail of the decay distribution.
In particular, due to a flux normalization, the results of the 
CERN-Heidelberg Collaboration [20] are insensitive to the interference effects;
therefore their data can be regarded as a determination of $R_{+-}^L$:
$R_{+-}^L=(5.29\pm 0.16)\times 10^{-6}$.
On the other hand, the attention of the CPLEAR Collaboration [21]
is concentrated on the interference region, 
so that their results can be taken to determine 
both the modulus $|\eta_{+-}|$ and the phase $\phi_{+-}$
of $\eta_{+-}$. Their most recent preliminary values for
these two quantities read [22]: $|\eta_{+-}|=(2.316 \pm 0.039)\times 10^{-3}$,
$\phi_{+-}=(43.5 \pm 0.8)^\circ$.

Other interesting observables, directly measured in experiments,
are the asymmetries associated with the decay into the final state $f$
of an initial $K^0$ as compared to the corresponding decay 
into $\bar f$ of an initial $\overline{K^0}$. 
Using the same notations as before, all these asymmetries have 
the general form
$$
A(t)={
{\rm Tr}\Big[\rho_{\bar{K^0}}(t){\cal O}_{\bar f}\Big]-
\Big[\rho_{K^0}(t){\cal O}_f\Big]\over
{\rm Tr}\Big[\rho_{\bar{K^0}}(t){\cal O}_{\bar f}\Big]+
\Big[\rho_{K^0}(t){\cal O}_f\Big]}\ ,\eqno(25)
$$
where $\rho_{\bar{K^0}}(t)$ and $\rho_{K^0}(t)$ are the solutions of
(3) with the initial conditions of having a pure $\overline{K^0}$
and pure $K^0$ at $t=0$, respectively. These correspond to the initial 
density matrices:
$$
\rho_{\bar{K^0}}=|{\overline{K^0}}\rangle\langle{\overline{K^0}}|\equiv
{1\over2}\left(\matrix{\phantom{-}1&-1\cr
                              -1&\phantom{-}1}\right)\ ,\eqno(26)
$$
and (21) above.

One of the simplest asymmetries that can be computed is the one that
involves $\pi^+\pi^-\pi^0$ in the final state. The operator ${\cal O}_{+-0}$
responsible for this decay can be obtained following steps similar to the ones
that led to the expressions of ${\cal O}_{+-}$ and ${\cal O}_{00}$ above.
Parameterizing the decay amplitudes as:
$$
{\cal A}(K_2\rightarrow \pi^+\pi^-\pi^0)=X_{+-0}\ ,\qquad
{\cal A}(K_1\rightarrow \pi^+\pi^-\pi^0)=Y_{+-0}\ 
{\cal A}(K_2\rightarrow \pi^+\pi^-\pi^0)\ ,\eqno(27)
$$
one easily finds: [15]
$$
{\cal O}_{+-0}=|X_{+-0}|^2\ \left[\matrix{|Y_{+-0}|^2 & Y_{+-0}^*\cr
                              Y_{+-0} & 1\cr}\right]\ .\eqno(28)
$$
Inserting this in (25) and using the solutions (11), one obtains, 
to first order in the small quantities, the following expression
for the asymmetry integrated over the entire Dalitz plot:
$$
\eqalign{
A_{+-0}(t)=&2\,{\cal R}e\left(\epsilon_S-{2 C\over\Delta\Gamma_-}\right)\cr
-&2\,e^{-\Delta\Gamma t/2}\Big[
{\cal R}e(\eta_{+-0})\, \cos(\Delta m\, t)
-{\cal I}m(\eta_{+-0})\, \sin(\Delta m\, t)
\Big]\ ,}\eqno(29)
$$
where 
$$
\eta_{+-0}\equiv\epsilon_S-{2 C\over\Delta\Gamma_-}+Y_{+-0}\ .\eqno(30)
$$
It is precisely the real and imaginary part of this parameter,
appearing in the interference
term of $A_{+-0}(t)$, that is usually fitted with the experimental data.
The best available determination of $\eta_{+-0}$ comes from the CPLEAR
Collaboration [22]. In particular, one finds:
${\cal R}e(\eta_{+-0})=(-4 \pm 8)\times 10^{-3}$. (The value of
${\cal I}m(\eta_{+-0})$ will not be needed for the considerations that
follow.)

Other interesting asymmetries involve 
decays into semileptonic states. The amplitudes for
the decay of a $K^0$ or a $\overline{K^0}$ state into $\pi^-\ell^+\nu$
and $\pi^+\ell^-\bar\nu$ are usually parametrized by three complex
constants $x$, $y$ and $z$  as follows:[23]
$$
\eqalignno{
&{\cal A}(K^0\rightarrow\pi^-\ell^+\nu)={\cal M} (1-y)\ , &(31a)\cr
&{\cal A}(\overline{K^0}\rightarrow\pi^+\ell^-\bar\nu)=
{\cal M}^* (1+y^*)\ , &(31b)\cr
&{\cal A}(K^0\rightarrow\pi^+\ell^-\bar\nu)= z\, 
{\cal A}(\overline{K^0}\rightarrow\pi^+\ell^-\bar\nu)\ , &(31c)\cr
&{\cal A}(\overline{K^0}\rightarrow\pi^-\ell^+\nu)=
x\, {\cal A}(K^0\rightarrow\pi^-\ell^+\nu)\ , &(31d) }
$$
where ${\cal M}$ is a common factor.
(In Ref.[23] $\bar x\equiv z^*$ is used instead of $z$.)
The $\Delta S=\Delta Q$ rule would forbid the decays
$K^0\rightarrow\pi^+\ell^-\bar\nu$ and 
$\overline{K^0}\rightarrow\pi^-\ell^+\nu$, so that the parameters $x$ and $z$
measure the violations of this rule. Instead, $CPT$-invariance would require
$y=\,0$.

From the parametrization in (31), one can deduce the decay amplitudes
from the states (1) of definite $CP$, and therefore the following
expressions for the two operators describing the semileptonic decays:
$$
\eqalignno{
&{\cal O}_{\ell^+}={|{\cal M}|^2\over2}\,
|1-y|^2\ \left[\matrix{|1+x|^2&(1+x^*)(1-x)\cr
                                (1+x)(1-x^*)&|1-x|^2\cr}\right]\ ,&(32a)\cr
&{\cal O}_{\ell^-}={|{\cal M}|^2\over2}\,
|1+y|^2\ \left[\matrix{|z+1|^2&(z^*+1)(z-1)\cr
                                (z+1)(z^*-1)&|z-1|^2\cr}\right]\ .&(32b)}
$$
The parameters $x$, $y$, and $z$ are expected to be very small, and the
available experimental determinations support this theoretical hypothesis. 
Therefore, in the following we shall treat these constants as the other small 
parameters in the theory and compute the various semileptonic observables
keeping only the first order terms in all these parameters.

With the help of (32), one can compute the following
two semileptonic asymmetries, for which
experimental data are available:
$$
\eqalignno{
&A_{\Delta m}(t)={
{\rm Tr}\Big[\rho_{\bar{K^0}}(t)\Big({\cal O}_{\ell^-} -
{\cal O}_{\ell^+}\Big)\Big] -
{\rm Tr}\Big[\rho_{K^0}(t)\Big({\cal O}_{\ell^-} -
{\cal O}_{\ell^+}\Big)\Big]
\over
{\rm Tr}\Big[\rho_{\bar{K^0}}(t)\Big({\cal O}_{\ell^-} +
{\cal O}_{\ell^+}\Big)\Big] +
{\rm Tr}\Big[\rho_{K^0}(t)\Big({\cal O}_{\ell^-} + 
{\cal O}_{\ell^+}\Big)\Big]}\, &(33a)\cr
&A_\ell(t)={
{\rm Tr}\Big[\rho_{\bar{K^0}}(t)\Big({\cal O}_{\ell^-} +
{\cal O}_{\ell^+}\Big)\Big] -
{\rm Tr}\Big[\rho_{K^0}(t)\Big({\cal O}_{\ell^-} + 
{\cal O}_{\ell^+}\Big)\Big]
\over
{\rm Tr}\Big[\rho_{\bar{K^0}}(t)\Big({\cal O}_{\ell^-} +
{\cal O}_{\ell^+}\Big)\Big] +
{\rm Tr}\Big[\rho_{K^0}(t)\Big({\cal O}_{\ell^-} + 
{\cal O}_{\ell^+}\Big)\Big]}\ .&(33b)}
$$
The first observable is sensible to the
$K^0\leftrightarrow \overline{K^0}$ oscillation frequency, while
the second one (also called $A_2(t)$) is the decay rate asymmetry
between $\overline{K^0}$ and $K^0$ in any semileptonic final state.
Using the solutions (11) of the kaon evolution equation, one finds the
following explicit expressions, valid to first order in all small
parameters:
$$
A_{\Delta m}(t)={N_{\Delta m}(t)\over D(t)}\ ,\qquad\quad
A_\ell(t)={N_\ell(t)\over D(t)}\ ,\eqno(34)
$$
where the numerators are given by 
$$
\eqalignno{
&N_{\Delta m}(t)=2\, e^{-\Gamma t}\bigg\{\left[1-(A-\gamma)t
\right]\, \cos(\Delta m\, t) \cr
&\hskip 4cm + \left[{\cal R}e\left({B\over \Delta m}\right)
-{\cal I}m(x+z)\right]\,
\sin(\Delta m\, t)\bigg\}\ ,&(35a)\cr
&N_\ell(t)=2\, {\cal R}e\left(\epsilon_L-{2 C^*\over\Delta\Gamma_-}\right)\
e^{-\gamma_S t} +
2\, {\cal R}e\left(\epsilon_S-{2 C\over\Delta\Gamma_-}\right)\
e^{-\gamma_L t}\cr 
&\hskip 1cm -2\, e^{-\Gamma t}\bigg\{
\left[{\cal R}e\left(\epsilon_L-{2 C^*\over\Delta\Gamma_-}\right)
+{\cal R}e\left(\epsilon_S-{2 C\over\Delta\Gamma_-}\right)\right]
\, \cos(\Delta m\, t)\cr 
&\hskip 1cm+\left[{\cal I}m\left(\epsilon_L-{2 C^*\over\Delta\Gamma_-}\right)
-{\cal I}m\left(\epsilon_S-{2 C\over\Delta\Gamma_-}\right)
-{\cal I}m(x-z)\right]\, \sin(\Delta m\, t)\bigg\}\ ,&(35b)}
$$
while the common denominator reads
$$
D(t)=\left[1-{2\gamma\over\Delta\Gamma}+{\cal R}e(x+z)\right]\, e^{-\gamma_S t} +
\left[1+{2\gamma\over\Delta\Gamma}-{\cal R}e(x+z)\right]\, e^{-\gamma_L t}
\ .\eqno(35c)
$$
(Notice that the expression in $(35a)$ is valid only for ``small'' 
times, {\it i.e.} as long as 
\hbox{$e^{-(A-\gamma)t}\sim 1-(A-\gamma)t$}.)

The asymmetry $A_{\Delta m}(t)$ can be used to determine the
parameters $a$ and $\alpha$, via the combinations $A=\alpha+a$
and ${\cal R}e(B)=\alpha-a$, once the
$\Delta S=\Delta Q$ rule is assumed 
and $\gamma$ has been fixed using other observables. 
On the contrary, notice that, within our approximation, 
the long-time limit of 
$A_\ell(t)/2$ gives the constant 
$2\,{\cal R}e(\epsilon_S-{2 C/\Delta\Gamma_-})$,
which is independent from $x$, $y$ and $z$. Therefore, this asymmetry 
allows a determination of this combination, 
independently from the $\Delta S=\Delta Q$ rule, and in alternative
to $A_{+-0}$, see (29).

Another useful observable involving semileptonic decays is
the so called $CP$ violating charge asymmetry:
$$
\delta(t)={
{\rm Tr}\Big[\rho(t)\left({\cal O}_{\ell^+}-{\cal O}_{\ell^-}\right)\Big]
\over
{\rm Tr}\Big[\rho(t)\left({\cal O}_{\ell^+}+{\cal O}_{\ell^-}\right)\Big]}
\ .\eqno(36)
$$
For a state which is initially a pure $K^0$,
it can be explicitly written as:
$$
\delta(t)={
\delta_S\, e^{-\gamma_S t}
+\delta_L\, e^{-\gamma_L t}
+2\, e^{-\Gamma t}\cos(\Delta m\,t)\over
e^{-\gamma_S t} + e^{-\gamma_L t} }\ ,\eqno(37)
$$
where
$$
\eqalignno{
&\delta_S=2\,{\cal R}e\left(\epsilon_S+{2C\over\Delta\Gamma_-}\right)
-{\cal R}e(2y+z-x)\ , &(38a)\cr
&\delta_L=2\, {\cal R}e\left(\epsilon_L+{2C^*\over\Delta\Gamma_-}\right)
-{\cal R}e(2y+x-z)\ ,&(38b)}
$$
are the charge asymmetries for the $K_S$ and $K_L$ states, respectively.
For sake of simplicity, in the expression (37) we have kept only
the dominant pieces in each term. Furthermore, notice that
the two expressions (38) can also be obtained by using 
directly the expressions (13) for $\rho_S$ and $\rho_L$.

Although not immediately connected with 
violations of a specific discrete symmetry, by comparing the
time-dependent asymmetry $\delta(t)$ with the experimental data
one can in principle measure at the same time the charge asymmetries for both 
the $K_S$ and $K_L$.
Unfortunately, the above expression for $\delta(t)$ has not yet been 
fitted with the experimental data, so that
one needs to look for separate determinations of $\delta_S$
and $\delta_L$. This last quantity is very well known from the fixed
target experiments, and for its value
one can take the world average given in [24]:
$\delta_L=(3.27 \pm 0.12)\times 10^{-3}$ 

On the contrary, no direct measurement of $\delta_S$ has been performed yet:
the available estimates on this quantity are deduced from other
observables, {\it e.g.} the two semileptonic asymmetries $A_T$ and $A_{CPT}$ 
obtained from the general formula (25) when the final state $f$ 
is $\pi^+\ell^-\bar\nu$ and $\pi^-\ell^+\nu$, respectively. 
The CPLEAR Collaboration has recently reported
a value for the combination $(A_T + A_{CPT})/2$, which in ordinary
quantum mechanics would give precisely an estimate for $\delta_S$ [22].
Although this is no longer strictly true when the generalized evolution
equation (3) is used to describe the decay of the
neutral kaon system (see [8]), 
in lacking of more complete and precise results
we shall take the CPLEAR value for $(A_T + A_{CPT})/2$
as an experimental determination of the $K_S$ charge asymmetry:
$\delta_S=(3.2 \pm 1.8)\times 10^{-3}$. 

We are now ready to derive estimates on some of the parameters
appearing in the dissipative piece $L[\rho]$ of the evolution equation (3).
In this analysis, we shall adopt some simplifying working assumptions.
First, we shall ignore the effects of $\epsilon^\prime$
in the two-pion decay rate (22), since one expects: 
$\epsilon^\prime/\epsilon_L\sim 10^{-4}$. Thus, the formulas presented
for the charged pion decay apply equally well to the neutron pion decay.
Also, we shall concentrate on possible $CPT$ violations in the time evolution
of the neutral kaon system as given by the dissipative part (9) and therefore
ignore the $CPT$ non preserving contributions in the various decay amplitudes;
this amounts to set the parameters $y$ and $B_0$, hence $Y_{+-}$, 
to zero. Furthermore,
we shall assume the validity of the $\Delta S=\Delta Q$ rule in the 
semileptonic decays; this choice is supported by theoretical arguments
that predict the parameters $x$ and $z$ to be neglegibly small.[19]

From the two and three-pion observables (24), (30) 
and the two semileptonic charge 
asymmetries (38),
one has:
$$
\eqalign{
&{\cal R}e\left(\epsilon_S+{2 C\over\Delta\Gamma_-}\right)={\delta_S\over2}
\ ,\cr
&{\cal R}e\left(\epsilon_S-{2 C\over\Delta\Gamma_-}\right)=
{\cal R}e(\eta_{+-0})\ ,}\qquad\quad
\eqalign{
&{\cal R}e\left(\epsilon_L+{2 C\over\Delta\Gamma_+}\right)={\delta_L\over2}
\ ,\cr
&{\cal R}e\left(\epsilon_L-{2 C\over\Delta\Gamma_+}\right)=
|\eta_{+-}|\, \cos(\phi_{+-})\ .}
\eqno(39)
$$
These relations allow extracting ${\cal R}e(\epsilon_S)$,
${\cal R}e(\epsilon_L)$ and
$$
\eqalignno{
&{c\over\Delta\Gamma}={1+\tau^2\over 16}\, \Big[\delta_S + \delta_L
-2\, \eta_{+-}\, \cos(\phi_{+-}) -2\, {\cal R}e(\eta_{+-0})\Big]\ , &(40a)\cr
&{\beta\over\Delta\Gamma}={1+\tau^2\over 16\tau}\, \Big[\delta_L - \delta_S
-2\, \eta_{+-}\, \cos(\phi_{+-}) +2\, {\cal R}e(\eta_{+-0})\Big]\ , &(40b)}
$$
where $\tau=2\Delta m/\Delta\Gamma$ is the tangent of the so called superweak
angle. From the previously quoted experimental values of the quantities
on the r.h.s. of (38), and taking into account 
that $\tau$ is very well known, [24]
$\tau= 0.9502 \pm 0.0034$, one obtains:
$$
c=(1.0 \pm 1.4)\times 10^{-17}\ \hbox{GeV}\ ,\qquad\qquad 
\beta=(-1.0 \pm 1.5)\times 10^{-17}\ \hbox{GeV}\ .
\eqno(41)$$

From these results, one can now estimate the value of the parameter
$\gamma$, using (23) and the determination of the $K_L$ two-pion decay rate
from the CERN-Heidelberg Collaboration:
$$
\eqalign{
{\gamma\over\Delta\Gamma}&=R^L_{+-}-|\eta_{+-}|^2
+{8\over 1+\tau^2}{\big(c^2+\beta^2\big)\over\Delta\Gamma^2}\cr
&+{4\over (1+\tau^2)\Delta\Gamma}\left[ {\cal R}e(\epsilon_L)\left(
(\tau^2-1)c-2\tau\beta\right)-
{\cal I}m(\epsilon_L)\left((\tau^2-1)\beta+2\tau c\right)\right]
\ ;}\eqno(42)
$$
${\cal I}m(\epsilon_L)$ can be obtained by taking the imaginary part
of (24): ${\cal I}m(\epsilon_L)=$\hfill\break 
$|\eta_{+-}|\, \sin(\phi_{+-}) + 2(\tau c-\beta)/[(1+\tau^2)\Delta\Gamma]$. 
In this way, one gets:
$$
\gamma=(0.1 \pm 30.2)\times 10^{-20}\ \hbox{GeV}\ .\eqno(43)
$$
Unfortunately, the errors propagation produces in this case a large uncertainty.

Estimates for the parameters $a$ and $\alpha$ can be obtained fitting
the asymmetry $(33a)$ with the available experimental data. Assuming again
$x=y=z=\,0$, taking $\gamma$ as in (43) and 
for $\Delta m$ the average of [24], we have performed
a preliminary $\chi^2$ fit of $A_{\Delta m}$ with the experimental data of the
CPLEAR Collaboration published in [25]. The result, rather independent from
the actual value of $\gamma$, gives: 
$a=(2.3 \pm 2.9)\times 10^{-17}\ \hbox{GeV}$ and
$\alpha=(2.1\pm 2.9)\times 10^{-17}\ \hbox{GeV}$. 
These values should be taken as indicative; a more accurate fit along the lines
discussed in [14] should provide more precise determinations.

The conclusion that can be drawn from this analysis is that 
the precision of the present experimental
data on the neutral kaon system is not high enough to allow a
meaningful test of the inequalities (10) and therefore of the
hypothesis of complete positivity. Nevertheless, as soon as more complete
and precise data become available, it will be possible to 
distinguish between simply positive 
and completely positive time-evolution maps in an actual experimental setup. 

In this respect, particularly promising are the planned experiments
on correlated neutral kaons at $\phi$-factories. In fact, entangled kaon
states might develop negative probabilities under simply positive time
evolutions. Since this is forbidden by complete positivity, very different
behaviours for entangled kaons are predicted in the two cases, with effects that
might be experimentally detectable [26].
In view of the wide use
of dynamical semigroups in the description of open systems [1, 2],
these checks of complete positivity might have consequences that
go beyond the specific situation of the neutral kaons.

\vfill\eject

\centerline{\bf REFERENCES}
\medskip

\item{1.} R. Alicki and K. Lendi, {\it Quantum Dynamical Semigroups and 
Applications}, Lect. Notes Phys. {\bf 286}, (Springer-Verlag, Berlin, 1987)
\smallskip
\item{2.} H. Spohn, Rev. Mod. Phys. {\bf 53} (1980) 569
\smallskip
\item{3.} V. Gorini, A. Kossakowski and E.C.G. Sudarshan,
J. Math. Phys. {\bf 17} (1976) 821
\smallskip
\item{4.} G. Lindblad,  Comm. Math. Phys. {\bf 48} (1976) 119
\smallskip 
\item{5.} E.B. Davies, {\it Quantum Theory of Open Systems}, (Academic Press,
New York, 1976)
\smallskip
\item{6.} V. Gorini, A. Frigerio, M. Verri, A. Kossakowski and
E.C.G. Surdarshan, Rep. Math. Phys. {\bf 13} (1978) 149 
\smallskip
\item{7.} F. Benatti and R. Floreanini, Phys. Lett. {\bf B389} (1996) 100
\smallskip
\item{8.} F. Benatti and R. Floreanini, Completely positive dynamical
maps and the neutral kaon system, Nucl. Phys. B, to appear
\smallskip
\item{9.} T.D. Lee and C.S. Wu, Ann. Rev. Nucl. Sci. {\bf 16} (1966)
511
\smallskip
\item{10.} J. Ellis, J.S. Hagelin, D.V. Nanopoulos and M. Srednicki,
Nucl Phys. {\bf B241} (1984) 381
\smallskip
\item{11.} J. Ellis, N.E. Mavromatos and D.V. Nanopoulos, Phys. Lett.
{\bf B293} (1992) 142
\smallskip
\item{12.} J. Ellis, N.E. Mavromatos and D.V. Nanopoulos, Int. J. Mod. Phys.
{\bf A11} (1996) 1489
\smallskip
\item{13.} J. Ellis, J.L. Lopez, N.E. Mavromatos and D.V. Nanopoulos, 
Phys. Rev. D {\bf 53} (1996) 3846
\smallskip
\item{14.} CPLEAR Collaboration, J. Ellis, N.E. Mavromatos and 
D.V. Nanopoulos, Phys. Lett. {\bf B364} (1995) 239
\smallskip
\item{15.} P. Huet and M.E. Peskin, Nucl. Phys. {\bf B434} (1995) 3
\smallskip
\item{16.} S. Hawking, Comm. Math. Phys. {\bf 87} (1983) 395
\smallskip
\item{17.} M. Srednicki, Nucl. Phys. {\bf B410} (1993) 143
\smallskip
\item{18.} F. Benatti and R. Floreanini, Testing complete positivity,
Trieste-preprint, 1996
\smallskip
\item{19.} L. Maiani, $CP$ and $CPT$ violation in neutral kaon decays,
in {\it The Second Da$\phi$ne Physics Handbook}, L. Maiani, G. Pancheri
and N. Paver, eds., (INFN, Frascati, 1995)
\smallskip
\item{20.} C. Geweniger {\it et al.}, Phys. Lett. {\bf B48} (1974) 487
\smallskip
\item{21.} R. Adler {\it et al.}, Phys. Lett. {\bf B286} (1992) 180
\smallskip
\item{22.} R. Adler {\it et al.}, Contribution of CPLEAR to the physics of the
neutral kaon system, CERN-PPE/96-189, 1996
\smallskip
\item{23.} N.W. Tanner and R.H. Dalitz, Ann. of Phys. {\bf 171} (1986)
463
\smallskip
\item{24.} Particle Data Group, Phys. Rev. D {\bf 54} (1996) 1
\smallskip
\item{25.} R. Adler {\it et al.}, Phys. Lett. {\bf B363} (1995) 237
\smallskip
\item{26.} F. Benatti and R. Floreanini, in preparation

\bye